\documentclass[prb, twocolumn, nofootinbib]{revtex4-2}

\usepackage{amssymb}
\usepackage{amsmath}
\usepackage{bbold}
\usepackage{mathrsfs} % mathscr
\usepackage{graphicx}
\usepackage{bm}
\usepackage{tabularx}
\usepackage[x11names]{xcolor}
\usepackage[unicode = true,colorlinks=true]{hyperref} %ACS doesn't want color.
\hypersetup{urlcolor = blue, linkcolor=blue }
\usepackage[normalem]{ulem}

\begin{document}
\title{Optical orientation of excitons and charged carriers in MAPbI$_3$ perovskite single crystals in the orthorhombic phase}

\author{Nataliia~E.~Kopteva$^{1}$, Dmitri~R.~Yakovlev$^{1}$, Ey\"up~Yalcin$^{1}$, Ilya~A.~Akimov$^{1}$, Mladen~Kotur$^{1}$, Bekir~Turedi$^{2,3}$, Dmitry~N.~Dirin$^{2,3}$, Maksym~V.~Kovalenko,$^{2,3}$ and Manfred~Bayer$^{1,4}$}

%all needed affiliations
\affiliation{$^{1}$Experimentelle Physik 2, Technische Universit\"at Dortmund, 44227 Dortmund, Germany}
\affiliation{$^{2}$Laboratory of Inorganic Chemistry, Department of Chemistry and Applied Biosciences,  ETH Z\"{u}rich, CH-8093 Z\"{u}rich, Switzerland}
\affiliation{$^{3}$Laboratory for Thin Films and Photovoltaics, Empa-Swiss Federal Laboratories for Materials Science and Technology, CH-8600 D\"{u}bendorf, Switzerland}
\affiliation{$^{4}$Research Center FEMS, Technische Universit\"at Dortmund, 44227 Dortmund, Germany}

\date{\today}
\makeatletter
\newenvironment{mywidetext}{%
  \par\ignorespaces
  \setbox\widetext@top\vbox{%
   \hb@xt@\hsize{%
    \leaders\hrule\hfil
    \vrule\@height6\p@
   }%
  }%
  \setbox\widetext@bot\hb@xt@\hsize{%
    \vrule\@depth6\p@
    \leaders\hrule\hfil
  }%
  \onecolumngrid
  \vskip10\p@
  \dimen@\ht\widetext@top\advance\dimen@\dp\widetext@top
  \cleaders\box\widetext@top\vskip\dimen@
  \vskip6\p@
  \prep@math@patch
}{%
  \par
  \vskip6\p@
  \setbox\widetext@bot\vbox{%
   \hb@xt@\hsize{\hfil\box\widetext@bot}%
  }%
  \dimen@\ht\widetext@bot\advance\dimen@\dp\widetext@bot
  %\cleaders\box\widetext@bot
  \vskip\dimen@
  \vskip8.5\p@
  \twocolumngrid\global\@ignoretrue
  \@endpetrue
}%
\makeatother

\begin{abstract}
Optical orientation of exciton and carrier spins by circularly polarized light is the basic phenomenon in the spin physics of semiconductors. Here, we investigate spin orientation in MAPbI$_3$ lead halide perovskite crystals at the cryogenic temperature of 1.6~K, where the material has an orthorhombic crystal structure. The recombination and spin dynamics of excitons and carriers are measured by time-resolved photoluminescence after circularly polarized excitation.  The optical orientation of excitons reaches 85\%, which persists within their lifetime of $15–80$~ps. This high orientation is maintained for excitation laser detunings from the exciton resonance to higher energies by up to 0.3~eV, then decreases and vanishes above 1.5~eV detuning. This indicates that the Dyakonov-Perel spin relaxation mechanism based on inversion symmetry breaking is inactive in MAPbI$_3$ crystals with orthorhombic symmetry. The optical orientation of localized and spatially-separated electrons and holes results in 40\% circular polarization of their emission. Their contributions can be identified from the complex spin beats dynamics in transverse magnetic field. The dynamics analysis gives values of the Land\'e $g$-factors of $|g_\text{V,e}| = 2.83$ for electrons and $|g_\text{V,h}| = 0.54$ for holes. Also, the magnetic-field-induced polarization of excitons and carriers is analyzed in magnetic fields up to 6~T, showing that their spin relaxation times are longer than their lifetimes. Namely, for the excitons, the spin relaxation time exceeds the lifetime by a factor of 6. We model the dynamics of optical orientation degree for cumulative contributions of excitons and carriers and show that the exciton recombination dynamics can control these dynamics. The polarized emission of excitons and localized carriers, produced by their polarization on Zeeman-split levels in magnetic fields, is modeled. Their dependence on the magnetic field is identical for very short spin relaxation times but becomes qualitatively different as the spin relaxation time approaches the carrier lifetime.
\end{abstract}

\maketitle

\section{Introduction}
\label{Introduction}

Lead halide perovskite semiconductors demonstrate remarkable electronic and optical properties, making them attractive for photovoltaic and optoelectronic applications~\cite{Vinattieri2021_book,Vardeny2022_book,Martinez2023_book}. They also have fascinating spin-dependent properties, which recently have been addressed by various magneto-optical techniques, without and with time-resolution~\cite{Giovanni2015,odenthal2017,belykh2019,Kopteva_gX_2024}. The perovskite band structure differs considerably from that of III-V and II-VI zinc-blende semiconductors, widely studied in solid state spin physics. Lead halide perovskites represent a novel prototype material for spin physics, providing access to novel experimental regimes and phenomena as well as to opportunities for exploring corresponding theoretical concepts. This potential is enhanced by the tunability of perovskite semiconductors, which allows adjustment of their bandgap across the visible spectral range from 1.5 to 3.2~eV and the possibility to obtain various crystal phases, comprising cubic, tetragonal, and orthorhombic structures. Furthermore, high-quality bulk single crystals of lead halide perovskites show sharp spectral lines in emission, absorption, and reflection, becoming as narrow as 1~meV at cryogenic temperatures. This enables the use of optical and magneto-optical techniques with high spectral resolution, allowing accurate measurements of spin-dependent parameters with high precision. MAPbI$_3$ is a hybrid organic-inorganic perovskite semiconductor that is representative for the whole material class. It has a tetragonal crystal structure at room temperature and an orthorhombic crystal structure below 160~K, and is instructive as reference for comparing its spin-dependent properties to the about cubic FA$_{0.9}$Cs$_{0.1}$PbI$_{2.8}$Br$_{0.2}$ crystals, which we studied recently in detail~\cite{kirstein2022am,XOO2024,COO2024}.      

Optical orientation of exciton and carrier spins is widely used in spin physics experiments. In this concept, circularly polarized photons generate spin-polarized excitons, electrons, and holes in semiconductors. The electronic band structure determines the maximum degree of optical orientation, which may reach 100\% in lead halide perovskites~\cite{XOO2024,XOO_Rashba}. For comparison, it is limited to 50\% in semiconductors with zinc-blende crystal structure, like GaAs and CdTe~\cite{OO_book}. Detection of the induced spin polarization and its dynamics can be done via  polarized photoluminescence (PL), Faraday (Kerr) rotation, or polarized differential transmission (absorption)~\cite{Spin_book_2017}. 

A high degree of optical orientation of the exciton PL of 85\% was recently reported for FA$_{0.9}$Cs$_{0.1}$PbI$_{2.8}$Br$_{0.2}$ crystals at 1.6~K temperature~\cite{XOO2024}. Remarkably, this high degree can be achieved even for excitation energies strongly detuned from the exciton resonance, indicating that the Dyakonov-Perel spin relaxation mechanism is absent in the about cubic FA$_{0.9}$Cs$_{0.1}$PbI$_{2.8}$Br$_{0.2}$ materials. This confirms that the spatial inversion symmetry is preserved in these materials, indicating the absence of spin-splittings in the conduction and valence bands, such as Rashba-Dresselhaus splittings~\cite{Kepenekian2015,Kepenekian2017}, which typically occur in crystals with a strong spin-orbit interaction and lack of an inversion center. Optical orientation of localized electrons and holes was also measured in FA$_{0.9}$Cs$_{0.1}$PbI$_{2.8}$Br$_{0.2}$ crystals. The degree of optical orientation reaches 60\% for short-living carriers~\cite{XOO2024} and decreases to a few percent for long-living carriers, measured using continuous-wave (cw) excitation~\cite{COO2024}. Localized electrons and holes couple to the nuclear spins, enabling dynamic nuclear polarization~\cite{COO2024,kirstein2022am}. Earlier studies on  optical orientation of carriers in perovskite semiconductors, detected through polarized PL using continuous-wave excitation, were limited to polycrystalline films: in a MAPbBr$_3$ film 3.1\% were detected at 10~K temperature~\cite{wang2019}, 2\%~\cite{Wang2018} and 8\%~\cite{wu2019} at 77~K, and in a MAPbI$_3$ film 0.15\% at 77~K~\cite{Wang2018}.

In this paper, we study the spin dynamics of excitons and localized carriers in MAPbI$_3$ thin crystals using the optical orientation technique at cryogenic temperatures. The spin polarization of excitons and carriers is detected by polarized photoluminescence, and time-resolved measurements are used to distinguish the different contributions. A high optical orientation degree of 85\% is found for excitons, which is stable with respect to excitation energy detuning from the exciton resonance. We measure the dynamics of spin polarization in longitudinal and transverse magnetic fields and conclude about the spin relaxation times of excitons and carriers. Model considerations are developed to analyze the spin dynamics for the case where the emission contributions from the two spin reservoirs overlap spectrally. 

The paper is organized as follows. In Section~\ref{Experimentals} information on perovskite samples and experimental techniques are given. The optical characterization of the studied MAPbI$_3$ thin crystals employing time-resolved PL and reflectivity is presented in Sec.~\ref{Optical properties}. Data on optical orientation of excitons and carriers and simulation of their spin dynamics are given in Secs.~\ref{OO excitons} and \ref{Modeling OO}. The dependence of exciton optical orientation on the detuning of excitation energy is examined in Sec.~\ref{OO detuning}. The dynamics of spin-oriented electrons and holes in a transverse magnetic field are considered in Sec.~\ref{OO carriers}. The exciton spin polarization in a magnetic field is measured and analyzed in Secs.~\ref{Exciton DCP} and \ref{Modeling DCP}. 

\section{Experimentals}
\label{Experimentals}

The studied {MA}PbI$_{3}$ single crystal was synthesized from PbI$_2$ and MAI perovskite precursors. The precursors were injected between two polytetrafluoroethylene coated glasses and slowly heated to 120$^\circ$C~\cite{Yang2022}. The sample has a square shape of about 2$\times$2~mm in the (001) crystallographic plane and a thickness of 30~$\mu$m (sample code M2-4). The {MA}PbI$_{3}$ single crystal has a tetragonal crystal structure at room temperature with an out-of-plane tetragonal $[001]$ axis. At cryogenic temperatures as used in our experimental studies, the crystal structure  transforms to orthorhombic. 

In all optical experiments, the geometry with the light wave vector $\textbf{k}\parallel [001]$ is used. We use a liquid helium cryostat with the temperature variable from 1.6~K to 300~K. At $T=1.6$~K, the sample is placed in superfluid helium, while at $4.2 - 85$\,K it is kept in helium vapor. A superconducting split-coil magnet generates magnetic fields up to 7~T, which can be applied either parallel to $\textbf{k}$ (denoted as $B_{\rm F}$ in the Faraday geometry) or perpendicular to it (denoted as $B_{\rm V}$ in the Voigt geometry).

Time-integrated photoluminescence (PL) and reflectivity spectra are measured with a 0.5\,m spectrometer equipped with a charge-coupled-device (CCD) camera. We use a halogen lamp as light source in the reflectivity measurements. For time-resolved PL measurements in the $\mu$s time range, a pulsed laser with photon energy of 3.5~eV, pulse duration of 4~ns, repetition rate of 5~kHz, and average excitation power of 12~$\mu$W is used. The signal is detected using an avalanche photodiode attached to the spectrometer and transferred to a time-of-flight card, providing a time resolution of 30~ns.

For measuring recombination and spin dynamics with a time resolution of 6~ps, a streak-camera (nominal time resolution of 2~ps) in combination with a 0.5\,m spectrometer equipped with a 300 grooves/mm diffraction grating is used. The photoluminescence is excited by a pulsed Coherent Chameleon Discovery laser combined with an APE HarmoniXX SHG unit. The laser pulses have 100~fs duration, 80~MHz repetition rate and are spectrally tunable from 1.7~eV (730~nm) to 3.1~eV (400~nm). We use circularly ($\sigma^+/\sigma^-$) polarized excitation to measure optical orientation and analyze the circularly polarized emission. Time-integrated PL spectra are obtained by integration of the PL dynamics over time.

\section{Results and discussion}
\label{Results}

\subsection{Optical properties} 
\label{Optical properties}

\begin{figure}[htb]
\begin{center}
\includegraphics[width = 7cm]{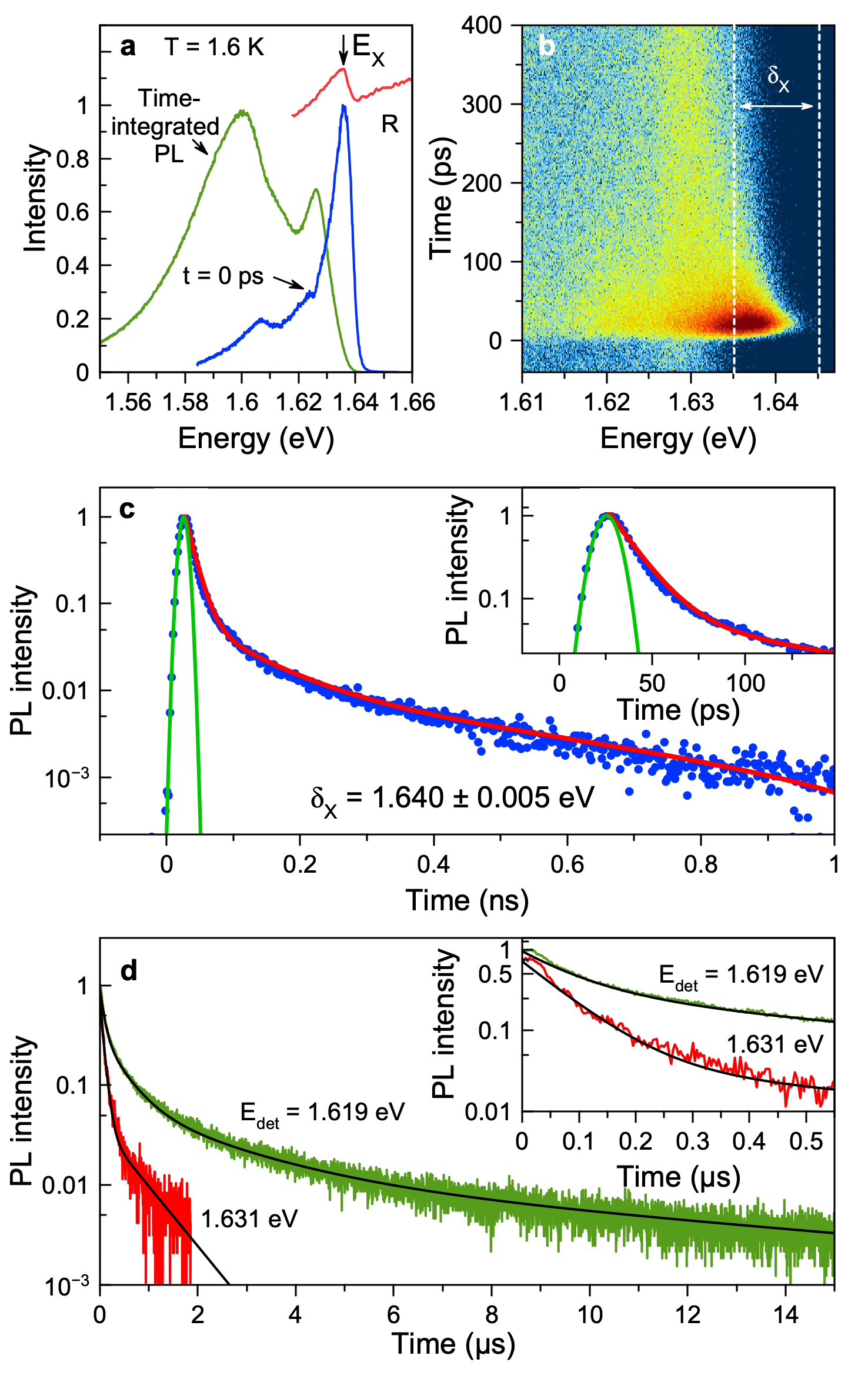}
\caption{\label{fig1} Recombination dynamics in the MAPbI$_{3}$ thin crystal measured at $T=1.6$~K.
(a) Reflectivity spectrum (red line) in the vicinity of the exciton resonance $E_\text{X}$ indicated by the arrow. The blue line is the PL spectrum right after the laser excitation, taken by time integration in the range of $0-5$~ps. The green line is the time-integrated PL spectrum. Laser parameters: $E_\text{exc} = 1.771$~eV and power density of  $P = 10$~mW/cm$^2$. The PL spectra are normalized to their maximal values. (b) Color-coded plot of time-resolved PL measured with the streak camera. (c) Recombination dynamics detected at the exciton resonance and integrated over the $\delta_\text{X}$ spectral range (circles). The red line is a three-exponential fit with decay times of $\tau_\text{X} = 15$~ps, $\tau_\text{X2} = 85$~ps, and $\tau_\text{R3} = 520$~ps. The green line is the pulse temporal profile with half-width on half maxima of 6~ps. The insert shows the initial parts of the PL dynamics. (d) Time-resolved PL measured at $E_\text{det} = 1.631$~eV (red) and 1.619~eV (green), each in a 15~$\mu$s range with $E_\text{exc} = 3.5$\,eV at the laser repetition rate of 5~kHz. The black lines are multi-exponential fits giving decay times collected in Table~\ref{tab:St1}. The insert shows the initial parts of the PL dynamics.}
\end{center}
\end{figure}

The optical properties of the studied MAPbI$_{3}$ crystal at $T = 1.6$~K are presented in Fig.~\ref{fig1}(a). In the reflectivity spectrum shown by the red line, a pronounced exciton resonance is observed at $E_\text{X} = 1.636$~eV. The exciton binding energy in bulk MAPbI$_{3}$ is 16~meV~\cite{galkowski2016}, therefore, the estimated band gap energy is $E_\text{g} = 1.652$~eV. The temporally- and spectrally-resolved PL measured with the streak camera is shown as color-coded plot in Fig.~\ref{fig1}(b). Right after laser excitation, the PL maximum is at the energy of 1.636~eV, corresponding to exciton emission; see also the blue spectrum in Fig.~\ref{fig1}(a). The exciton emission shows a fast decay, after which the PL maximum shifts to lower energies, ultimately approaching 1.627~eV. This is evidenced by the time-integrated spectrum (green line) in Fig.~\ref{fig1}(a). Also, the low energy band with the maximum at 1.600~eV becomes most intense in the time-integrated spectrum. 

The PL dynamics spectrally-integrated in the range $\delta_\text{X} = 1.640\pm0.005$~eV, i.e. in the vicinity of the exciton resonance, are shown in Fig.~\ref{fig1}(c). In the temporal range up to 1~ns, the dynamics have three-exponential decay with two fast decay times $\tau_\text{X} = 15$~ps and $\tau_{\text{X}2} = 85$~ps  assigned to the exciton recombination. The long component with $\tau_{\text{R}3} = 520$~ps is due to the recombination of spatially separated, localized electrons and holes, as is typical for bulk lead halide perovskite semiconductors for which the recombination lines of electron-hole pairs and excitons are spectrally overlapping~\cite{XOO2024,COO2024,kirstein2022mapi,deQuilettes2019_si,herz2017}. As shown here, their contributions can be separated by time-resolved techniques and polarized PL in magneto-optical experiments~\cite{COO2024}.   

The Stokes-shifted PL shows a recombination dynamics lasting longer than the 12~ns of the repetition period of the mode-locked laser, operating at 80~MHz frequency. Therefore, we use the laser operating at 5~kHz repetition frequency in combination with a time-of-flight detection electronics. The dynamics measured at 1.631~eV and 1.619~eV in the temporal range up to 15~$\mu$s are shown in Fig.~\ref{fig1}(d). The lower time resolution in these experiments of 30~ns does not allow us to resolve any fast dynamics, but still one can see from the fits and the resulting parameters given in Table~\ref{tab:St1} that the dynamics are multi-exponential. Namely, they contain two components at 1.631~eV energy and four components at 1.619~eV. This indicates the presence of several recombination processes. The long recombination dynamics can be attributed to the recombination of electrons and holes that are localized at different crystal sites, with significant variation in their spatial separation~\cite{kirstein2022mapi,deQuilettes2019_si,herz2017}. Additional processes such as polaron formation~\cite{deQuilettes2019_si}, carrier trapping and detrapping process~\cite{Chirvony2018}, as well as in-depth carrier diffusion~\cite{Bercegol2018} may also contribute. However, estimating their impact on the measured signal is beyond the scope of our study.    

\begin{table*}[htb]
\caption{Recombination times in the MAPbI$_{3}$ crystal evaluated from the PL dynamics given in Fig.~\ref{fig1}(d) using a multi-exponential fit of the form $I(t)=\sum_{i}A_{i} \exp(-t/\tau_{i})$. We normalize the intensities of the PL components so that $\sum_{i}A_{i} =1$. }
\label{tab:St1}
\begin{center}
\begin{tabular*}
{0.99\textwidth}{@{\extracolsep{\fill}} |>{\centering\arraybackslash} m{0.1\textwidth} |>{\centering\arraybackslash} m{0.1\textwidth}|>{\centering\arraybackslash} m{0.1\textwidth}|>{\centering\arraybackslash} m{0.1\textwidth}|>{\centering\arraybackslash} m{0.1\textwidth}|>{\centering\arraybackslash} m{0.1\textwidth}|>{\centering\arraybackslash} m{0.1\textwidth}|>{\centering\arraybackslash} m{0.1\textwidth}|>{\centering\arraybackslash} m{0.1\textwidth}|}
\hline
$E_\text{det}$\,(eV) & $A_1$ & $\tau_1$\,(ns) & $A_2$ & $\tau_2$\,(ns) & $A_3$ & $\tau_3$\,($\mu$s)& $A_4$ & $\tau_4$\,($\mu$s)\\
\hline
1.631 & 0.94& 80 & 0.06 &730 & - & - & - & - \\
\hline
1.619 &0.65& 85 &0.27& 430 &0.06& 1.9 &0.02& 11 \\
\hline
\end{tabular*}
\end{center}
\end{table*}

\subsection{Optical orientation of excitons and carriers}
\label{OO excitons}

For measuring the optical orientation of excitons and carriers, we use $\sigma^+$ circularly polarized laser excitation and detect the PL both in $\sigma^+$ and $\sigma^-$ polarization. The optical orientation degree is evaluated from
\begin{equation}
\label{eq1}
P_{\rm{oo}} = \frac{I^{++} - I^{+-}}{I^{++} + I^{+-}}.
\end{equation}
Here, $I^{++}$ and $I^{+-}$ are the $\sigma^+$ and $\sigma^-$ polarized PL intensities for $\sigma^+$ polarized excitation. In time-resolved experiments for analyzing the $P_{\rm{oo}}$ dynamics, the dynamics of the respective PL intensities $I^{++}(t)$ and $I^{+-}(t)$ are used. 

\begin{figure}[hbt]
\begin{center}
\includegraphics[width = 7cm]{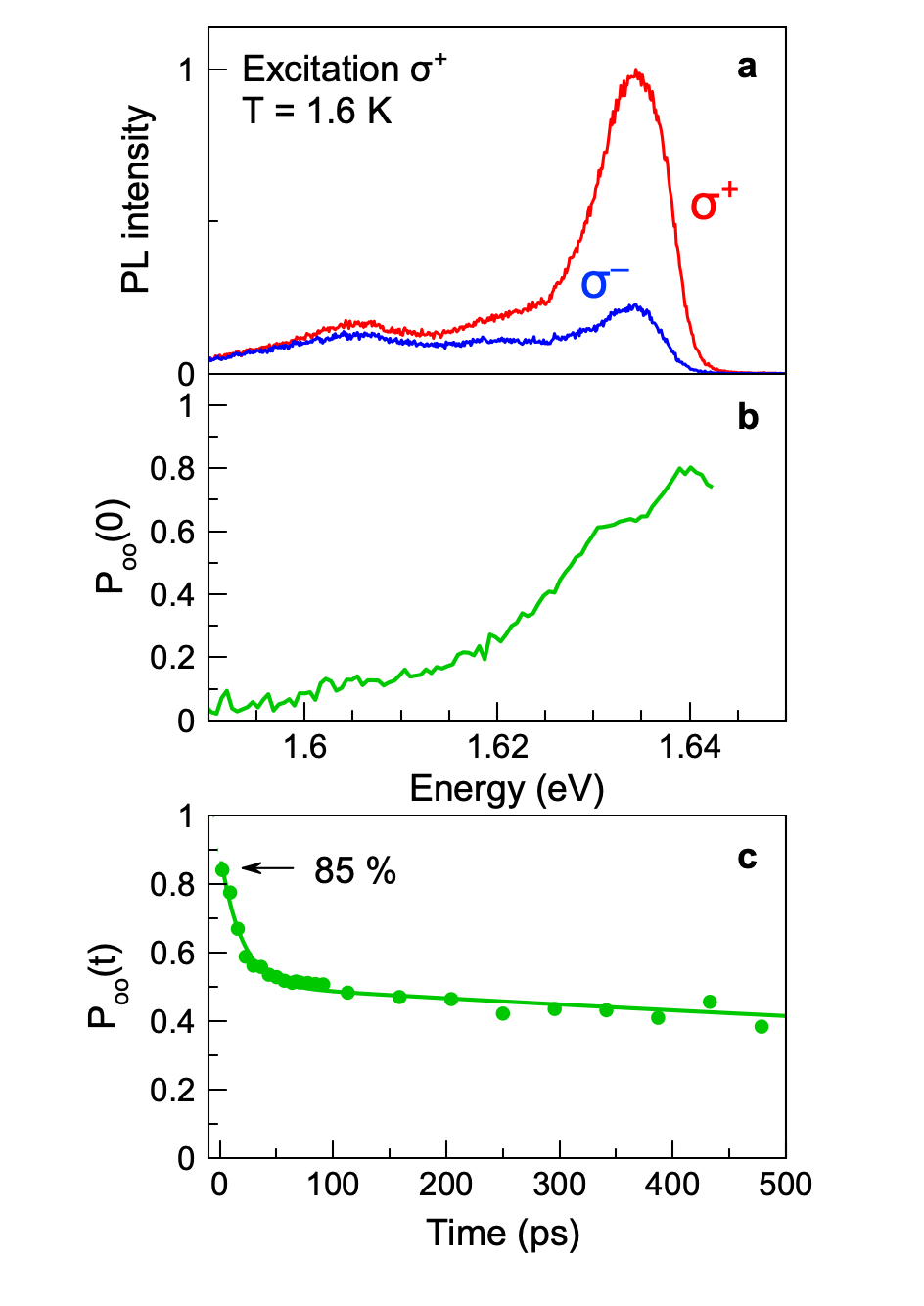}
\caption{\label{OO_intro} Optical orientation of excitons measured at $T = 1.6$~K for $\sigma^{+}$ excitation at $E_\text{exc} = 1.698$~eV using $P = 10$~mW/cm$^2$. (a) PL spectra integrated over the time range of $0-5$~ps after the laser pulse excitation and measured in  $\sigma^{+}$ (red) and $\sigma^{-}$ (blue) polarization.  (b) Spectral dependence of the optical orientation degree for the spectra from panel (a). (c) Dynamics of $P_\text{oo}(t)$, detected at $E_\text{det} = 1.640 \pm 0.005$~eV. The line is a fit giving the parameters collected in Table~\ref{tab:St2}.}
\end{center}
\end{figure}

Figure~\ref{OO_intro}(a) shows PL spectra right after pulsed excitation (time integrated in a 5~ps range), measured in $\sigma^+$ (red) and $\sigma^-$ (blue) polarization, after excitation with $\sigma^+$ polarized light. The stronger intensity of the $\sigma^+$ polarized emission highlights the effect of optical orientation.  The corresponding spectral dependence of $P_\text{oo}(0)$ is shown in Fig.~\ref{OO_intro}(b). The dependence has a pronounced maximum of $P_\text{oo}(0) = 0.82$ at the high energy flank of the exciton resonance at $E_\text{det} = 1.640$~eV.

The dynamics of optical orientation measured at $E_\text{det} = 1.640$~eV are shown in Fig.~\ref{OO_intro}(c). Right after the excitation pulse, the optical orientation $P_\text{oo}$ reaches 0.85, i.e.  85\%. Then, within 20~ps, it decreases to 0.50 (50\%), where it converts to a much slower decrease with a decay time of about 2.5~ns, evaluated from a biexponential fit. All recombination and spin dynamics parameters are collected in Table~\ref{tab:St2}. We show below in Sec.~\ref{Modeling OO} that in a case when several processes are involved, the optical orientation dynamics can be contributed not only by spin relaxation times, but also by recombination times. In our MAPbI$_3$ crystal, the initial fast decrease of $P_\text{oo}(t)$ is determined by the exciton lifetime, while the longer dynamics are given by the spin relaxation of localized electrons and holes. 

\subsection{Modeling of optical orientation dynamics contributed by excitons and carriers}
\label{Modeling OO} 
Let us start by considering the dynamics of optical orientation in a scenario in which only a single spin system, either the excitons or the electron-hole pairs, is involved. The dynamics of the exciton (or electron-hole pair) population can be described by:
\begin{equation}
\label{population}
I(t) =  I^{++}(t )+ I^{+-}(t) = I_0 \exp({-t/\tau}),
\end{equation}
where $I(t)$ is the PL intensity, and $I_0$ is the initial population just after the laser pulse action. $\tau$ is the recombination time. Note, that the experimental value $I(t) =  I^{++}(t) + I^{+-}(t)$ is the denominator in Eq.~\eqref{eq1}. The numerator in Eq.~\eqref{eq1} is the spin density:
\begin{equation}
\label{spin_density}
I^{++}(t ) - I^{+-}(t) = S_0 \exp({-t/T_\text{s}}),
\end{equation}
where $S_0$ is the spin density right after pulse action. $T_\text{s}$ is the spin lifetime, contributed by the recombination time $\tau$ and the spin relaxation time $\tau_s$: $T_\text{s}^{-1} = \tau^{-1} + \tau_\text{s}^{-1}$.

The resulting optical orientation degree, calculated by Eqs.~\eqref{eq1}-\eqref{spin_density}, decays with the spin relaxation time $\tau_\text{s}$:
\begin{equation}
 P_\text{oo}(t) = P_\text{oo}(0) \exp({-t/\tau_\text{s}}) \,.
 \label{eqOO_dyn}
\end{equation}        
Here, $P_\text{oo}(0)$ is the initial optical orientation degree, right after the pulse action. In experiment, it accounts for partial depolarization of the excitons and carriers during their energy relaxation. In this case, it can be considered as initial optical orientation degree for cold excitons and carriers just reaching the band edges.

\begin{figure}[t]
\begin{center}
\includegraphics[width = 8.5cm]{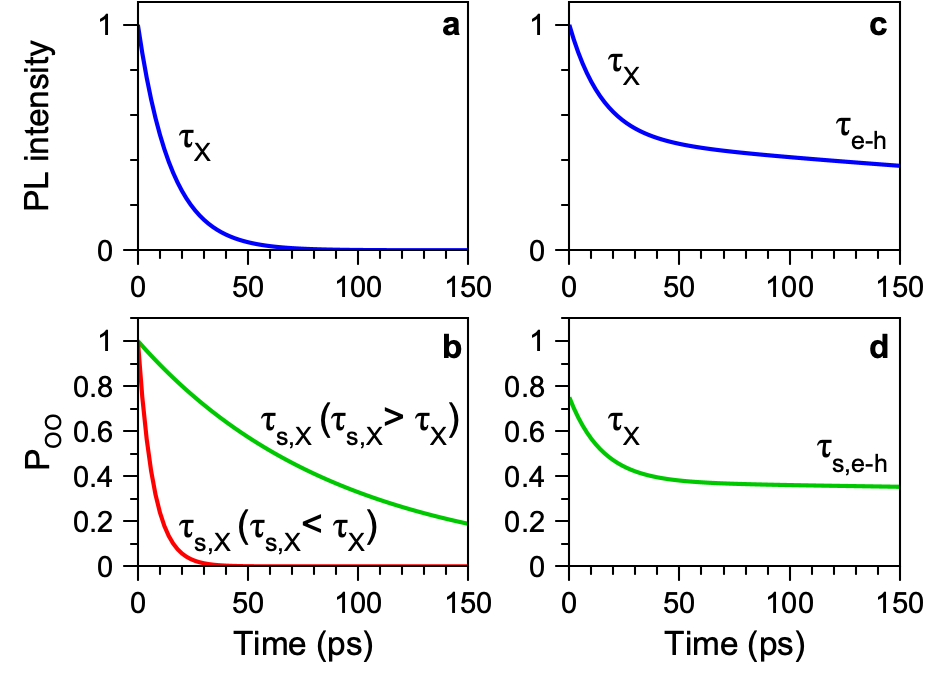}
\caption{\label{Sim_dyn} Simulation of exciton and electron-hole pair recombination and spin dynamics. (a) Exciton recombination dynamics characterized by the exciton lifetime $\tau_\text{X} = 15$\,ps. (b) Optical orientation dynamics for a long spin relaxation time $\tau_\text{s,X} > \tau_\text{X}$,  $\tau_\text{s,X} = 90$\,ps (green) and a short spin relaxation time $\tau_\text{s,X} < \tau_\text{X}$,  $\tau_\text{s,X} = 7$\,ps (red). (c) Dynamics involving both excitons and electron-hole pairs recombination with characteristic decay times $\tau_\text{X} = 15$\,ps and $\tau_\text{e-h} = 520$\,ps. (d) Green line gives the spin dynamics of the two spin systems. $\tau_\text{s,e-h} = 2.5$\,ns.}
\end{center}
\end{figure}

The case of a single spin system is illustrated in Figs.~\ref{Sim_dyn}(a,b) using excitons with $\tau_\text{X} = 15$~ps as example. As shown in Fig.~\ref{Sim_dyn}(a), the PL dynamics is determined by $\tau_\text{X}$. Regardless of whether the spin relaxation time is shorter ($\tau_\text{s,X} = 7$~ps, red line) or longer ($\tau_\text{s,X} = 90$~ps, green line) than $\tau_\text{X}$, the dynamics of $P_\text{oo}(t)$ are governed solely by $\tau_\text{s,X}$, as shown in Fig.~\ref{Sim_dyn}(b). Thus, for a single spin system, measuring the optical orientation dynamics directly reveals the spin relaxation dynamics.

For the case of two spin systems contributing to the spin dynamics at the same energy, we consider excitons and electron-hole pairs with recombination times fulfilling the relation $\tau_\text{X} \ll \tau_\text{e-h}$. As an example, we assume $\tau_\text{X}=15$~ps,  $\tau_\text{e-h}=520$~ps, and $I_\text{0,X}=I_\text{0,e-h}$. The PL dynamics described by   
\begin{equation}
I(t) = I_\text{0,X} \exp({-t/\tau_\text{X}}) + I_\text{0,e-h} \exp({-t/\tau_\text{e-h}}) 
 \label{eqRec_dyn} 
\end{equation}
are given in Fig.~\ref{Sim_dyn}(c), showing a biexponential decay with $\tau_\text{X}$ and $\tau_\text{e-h}$. 

The optical orientation dynamics, in this case, are described by: 
\begin{equation}
 \label{eqOO_dyn}
\begin{aligned}
 P_\text{oo}(t) = \frac{S_{0,\text{X}} \exp({-t/T_\text{s,X}}) + S_{0,\text{e-h}} \exp({-t/T_\text{s,e-h}}) }{I_\text{0,X} \exp({-t/\tau_\text{X}}) + I_\text{0,e-h} \exp({-t/\tau_\text{e-h}})}.
\end{aligned}
\end{equation}

The result of this modeling is shown in Fig.~\ref{Sim_dyn}(d) by the green line for the case $\tau_\text{s,X} \gg \tau_\text{X}$. The experimentally measured optical orientation dynamics shown in Fig.~\ref{OO_intro}(c) is in agreement with the scenario of two spin systems contributing to the spin dynamics. Initially, the $P_\text{oo}(t)$ signal is dominated by the excitons with the maximal value of $P_\text{oo,X}(0) = 0.85$. Once the excitons  recombine, the PL signal becomes dominated by long-lived carriers with $P_\text{oo,e-h}(0) = 0.40$. The decay of $P_\text{oo}$ from 0.85 to 0.40 is thus governed by the exciton lifetime $\tau_{\rm X}$. 

\begin{table*}[htb]
\caption{Exciton and electron-hole recombination and spin parameters in MAPbI$_{3}$ measured at $T = 1.6$\,K.}
\label{tab:St2}
\begin{center}
\begin{tabular*}
{0.67\textwidth}{@{\extracolsep{\fill}} |>{\centering\arraybackslash} m{0.1\textwidth} |>{\centering\arraybackslash} m{0.18\textwidth}|>{\centering\arraybackslash} m{0.2\textwidth}|>{\centering\arraybackslash} m{0.15\textwidth}|}
\hline
 & Recombination time & Spin relaxation time & $P_\text{oo}(0)$\\
\hline
Exciton & $\tau_\text{X} = 15$~ps & $\tau_\text{s,X} = 90$~ps (DCP) & 0.85 \\
\hline
Carriers &  $\tau_\text{e-h} = 520$~ps & $\tau_\text{s,e-h} = 2.5$~ns & 0.40\\
\hline
\end{tabular*}
\end{center}
\end{table*}

\subsection{Optical orientation of excitons generated with large kinetic energies}
\label{OO detuning}

Investigation of the degree of optical orientation on the detuning of the laser excitation energy from the exciton resonance gives information on the exciton and carrier spin dynamics for large kinetic energies and also on the matrix elements of the optical transitions between the involved electronic bands at these energies. Recently, we studied this for bulk FA$_{0.9}$Cs$_{0.1}$PbI$_{2.8}$Br$_{0.2}$ crystals with about cubic crystal structure~\cite{XOO2024} and it is instructive to extend these investigations to MAPbI$_3$ crystals with orthorhombic crystal structure.    

\begin{figure*}[htb]
\begin{center}
\includegraphics[width = 14cm]{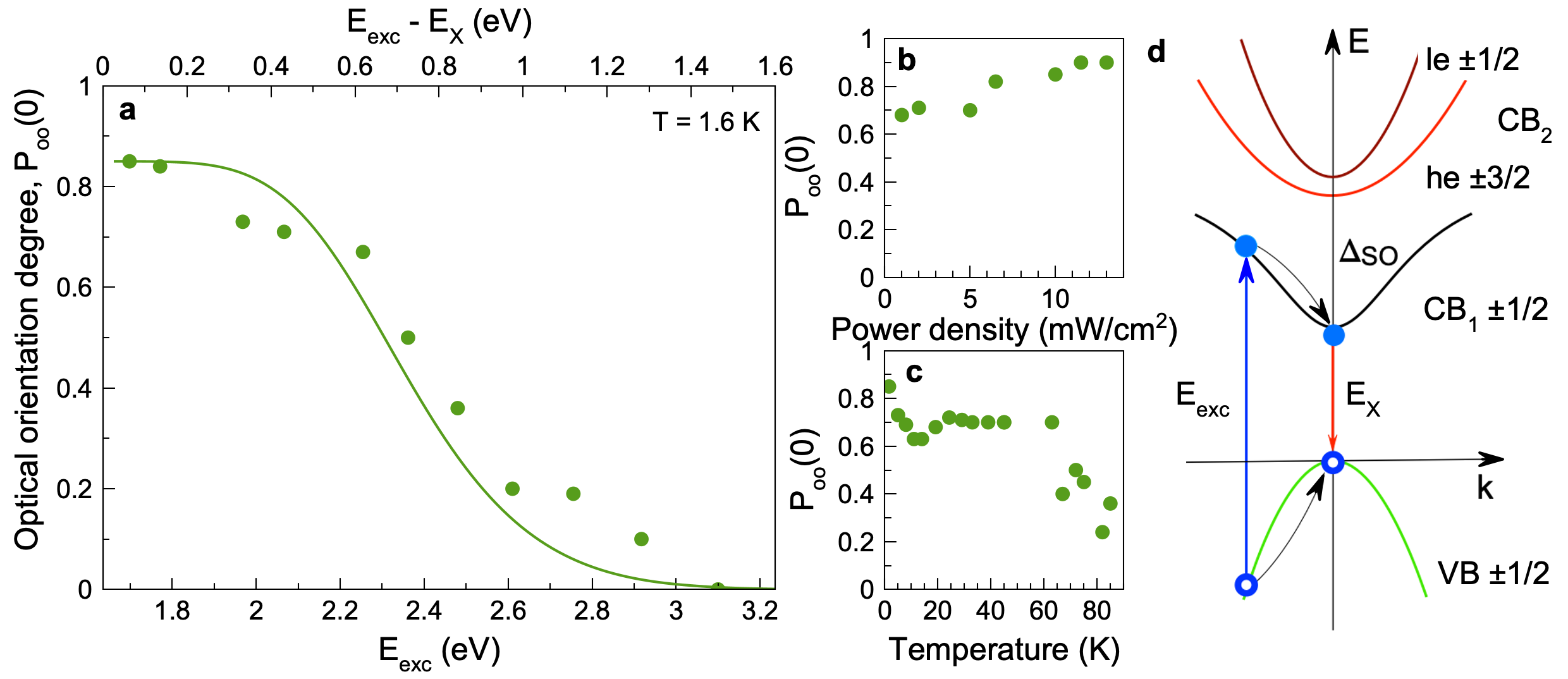}
%includegraphics[trim=0mm 0mm 0mm 0mm, clip, width=2.05\columnwidth]{Fig02.pdf}
\caption{\label{OO} Optical orientation of excitons detected at the exciton resonance at 1.640~eV energy. (a) $P_\text{oo} (0)$ dependence on the laser excitation energy $E_\text{exc}$ for $\sigma^+$ polarized excitation with $P = 10$~mW/cm$^2$ at $T = 1.6$~K. The upper axis gives the photon energy detuning from the exciton resonance $E_\text{exc} - E_\text{X}$. The line shows the theoretical dependence from Ref.~\cite{XOO2024}, accounting for the Elliott-Yafet spin relaxation due to interaction with longitudinal optical phonons. (b) $P_\text{oo} (0)$ as function of the excitation density measured at $T = 1.6$~K. (c) Temperature dependence of $P_\text{oo} (0)$ for $P = 10$~mW/cm$^2$ excitation density. In panels (b,c), the excitation photon energy $E_\text{exc} = 1.698$~eV. (d) Scheme of the lead halide perovskite band structure with orthorhombic symmetry. The valence (VB) and conduction (CB$_1$) bands are contributed by electron and hole states with spins $\pm1/2$. The upper conduction band (CB$_2$) is shifted by the spin-orbit splitting $\Delta_\text{SO}$ to higher energies and consists of light (le) and heavy (he) electron bands.}
\end{center}
\end{figure*}

Figure~\ref{OO}(a) shows the $P_\text{oo}(0)$ dependence on the excitation energy measured at the exciton resonance energy of 1.640~eV. The optical orientation degree weakly decreases from 0.85 to 0.70 when the excitation energy is tuned from 1.70~eV to 2.30~eV, corresponding to 0.60~eV detuning. For larger detunings, it decreases faster and approaches zero at $E_\text{exc} = 3.1$~eV, corresponding 1.5~eV detuning. A similar dependence was measured for FA$_{0.9}$Cs$_{0.1}$PbI$_{2.8}$Br$_{0.2}$ and discussed in detail in Ref.~\onlinecite{XOO2024}. We show in Fig.~\ref{OO}(a) the theoretical dependence from this paper by the green line, calculated by accounting for the modification of the optical matrix elements for large detunings and the Elliott-Yafet spin relaxation mechanism due to interaction with longitudinal optical phonons. This dependence is in good agreement with the experimental data in FA$_{0.9}$Cs$_{0.1}$PbI$_{2.8}$Br$_{0.2}$, and also closely follows the experimental data for the MAPbI$_3$ crystal. This allows us to conclude that the Dyakonov-Perel spin relaxation mechanism is absent in MAPbI$_3$ and that the reduction of the crystal symmetry from about cubic to orthorhombic does not provide a breaking of the spatial inversion symmetry. Note that  breaking of the spatial inversion symmetry in combination with strong spin-orbit interaction is the perquisite of Rashba-Dresselhaus spin splittings, facilitating the Dyakonov-Perel spin relaxation mechanism~\cite{Kepenekian2015,Kepenekian2017}. 

In our experiments, we measure the optical orientation of cold excitons after relaxation into their 1S state, while these excitons are photoexcited with large kinetic energy. Two energy relaxation scenarios can be suggested. For the first one, the exciton photogenerated with a small wave vector (corresponding to the light wavevector in the medium at this energy) at the upper (photon) polariton branch is scattered by phonons with a large wavevector to the lower (mechanical) polariton branch. Then, it relaxes through emission of optical and acoustic phonons to this lowest 1S state. This exciton involves the same electron-hole pair that was created by photogeneration, which can be confirmed by carrier spin correlation effects, that can be identified in exciton optical alignment or in exciton spin beats detected in linear polarization. Previously, we observed such a behavior in FA$_{0.9}$Cs$_{0.1}$PbI$_{2.8}$Br$_{0.2}$ crystals~\cite{XOO2024} and similar features are found for the studied MAPbI$_3$ crystals, which will be described elsewhere. Therefore, we conclude that a fraction of the excitons follows this energy relaxation scenario.       

For the second scenario, the photogenerated exciton, whose kinetic energy greatly exceeds its binding energy, is dissociated into an electron and a hole with large and opposite to each other momenta. The electron and hole individually relax toward the edges of the bands (Fig.~\ref{OO}(d)), where they can be either localized and contribute to the emission of spatially-separated electron-hole pairs or form excitons. However, these excitons are formed from nongeminate electron-hole pairs, i.e., the contributing carriers are photogenerated by different photons. Such excitons can retain a memory about the circular polarization of the exciting laser, but do not show any spin alignment. Optically oriented carriers are generated in this scenario, and we suggest that a fraction of the optically oriented excitons are also provided in that way. 

Figure~\ref{OO}(b) presents the $P_\text{oo}(0)$ dependence on the excitation density, varied from 1 to 13~mW/cm$^2$. The optical orientation of excitons is weakly sensitive to the excitation density in this range of powers. The maximal used excitation density of 13~mW/cm$^2$ corresponds to an exciton density of about  $1.3 \times 10^{13}$~cm$^{-3}$, which is relatively small for effects related to exciton-exciton interaction to become important.

\begin{figure*}[htb]
\begin{center}
\includegraphics[width = 12cm]{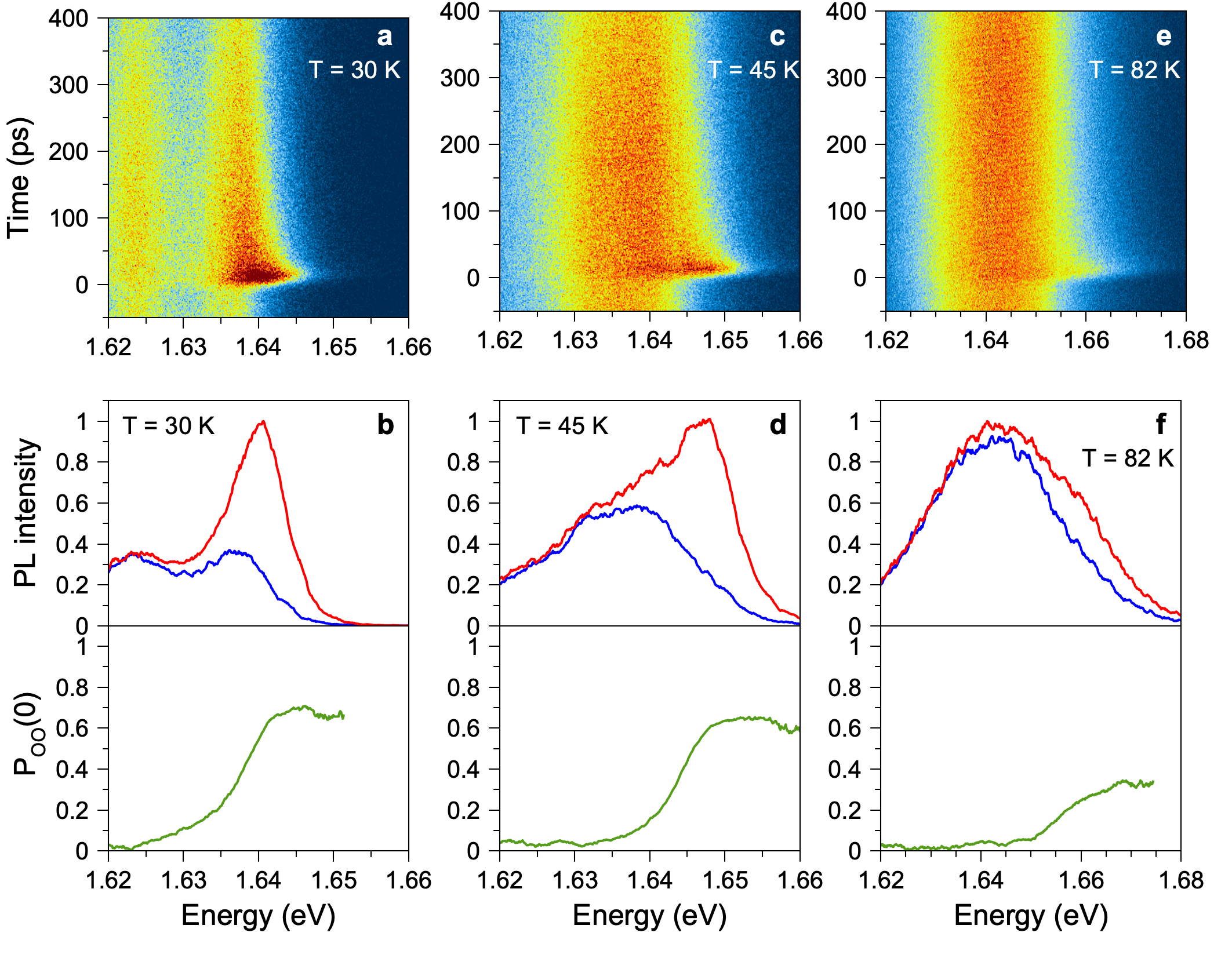}
%includegraphics[trim=0mm 0mm 0mm 0mm, clip, width=2.05\columnwidth]{Fig02.pdf}
\caption{\label{OO_TD} Temperature dependence of the PL dynamics and optical orientation degree $P_\text{oo}(0)$. The PL dynamics are shown for: (a) $T = 30$\,K, (c) $45$\,K, and (e) $82$~K. (b,d,f) PL spectra integrated over the time range of $0-5$~ps after laser pulse excitation and measured in $\sigma^{+}$ (red) and $\sigma^{-}$ (blue) polarization, respectively, at these temperatures. Below, the spectral dependences of the optical orientation degree calculated from these spectra.}
\end{center}
\end{figure*}

The optical orientation of excitons is stable with respect to temperature changes from $1.6-60$~K and decreases to 0.2 at 85~K, as shown in Fig.~\ref{OO}(c). A detailed picture of the PL dynamics and the spectral dependence of $P_\text{oo}(0)$ at the three temperatures of 30~K, 45~K, and 82~K  is given in Fig.~\ref{OO_TD}. 

\subsection{Optically oriented electrons and holes in transverse magnetic field}
\label{OO carriers}

\begin{figure*}[hbt]
\begin{center}
\includegraphics[width = 16cm]{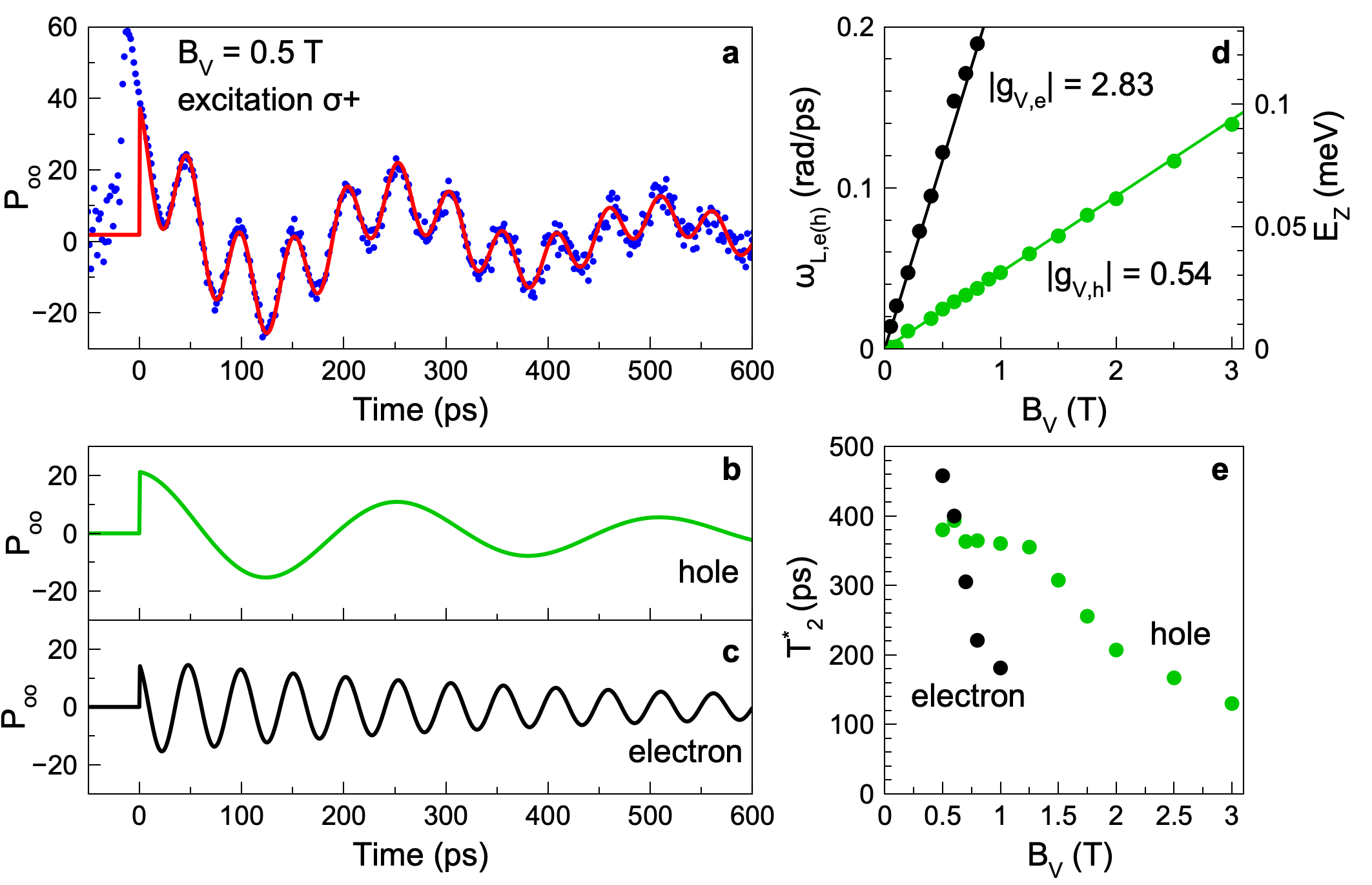}
\caption{\label{QB} (a) Dynamics of the optical orientation of electrons and holes measured in the Voigt magnetic field of $B_\text{V} = 0.5$\,T for $\sigma^+$ polarized excitation (symbols) at $T = 1.6$~K. $E_\text{exc} = 1.771$\,eV with $P = 10$\,mW/cm$^2$. $E_\text{det} = 1.640$~eV. Red line is a fit with Eq.~\eqref{Poo_beats1}, accounting for the hole and electron contributions, shown separately in panels (b) and (c). (d) Dependences of the Larmor precession frequency of the electrons (black circles) and the holes (green circles) on the Voigt field $B_\text{V}$. Linear fits give $|g_\text{V,e}| = 2.83$ and $|g_\text{V,h}| = 0.54$. (e) Electron (black) and hole (green) spin dephasing time ($T_2^*$) as function of $B_\text{V}$.}
\end{center}
\end{figure*}

Figure~\ref{QB}(a) shows the dynamics of $P_{\rm oo}(t)$ measured at the exciton energy ($E_\text{det} = 1.640$\,eV) in a magnetic field applied perpendicular to the light $k$-vector (Voigt geometry) of $B_\text{V} = 0.5$~T. A complex spin-beat pattern is observed, comprising a slow decay over a temporal range of 600 ps. The decay time is significantly longer than the exciton lifetime, which indicates that the signal arises from coherent spin precession of spatially separated, localized electrons and holes. 

We fit the optical orientation dynamics using the form~\cite{kocher98_si}:
\begin{equation}
\label{Poo_beats1}
P_\text{oo}(t) = \frac{P_\text{oo,e}+P_\text{oo,h}}{1+P_\text{oo,e}P_\text{oo,h}},
\end{equation}
with $P_\text{oo,e}$ and $P_\text{oo,h}$ being the optical orientation degree of electrons and holes:
\begin{equation}
\label{Poo_beats2}
P_{\rm{oo,e(h)}}(t) = P_{\rm oo,e(h)}(0) \cos(\omega_{\rm L,e(h)}t)\exp(-t/T^*_{\rm 2,e(h)}). 
\end{equation}
Here $P_{\rm oo,e(h)}(0)$ is the spin polarization degree of the electrons (holes) at zero time delay, $\omega_{\rm L,e(h)}$ is the electron (hole) Larmor precession frequency. $T^*_{\rm 2,e(h)}$ is the spin dephasing time of electrons (holes). We fit the dynamics in Fig.~\ref{QB}(a) with Eqs.~\eqref{Poo_beats1}-\eqref{Poo_beats2}. The separate dynamics of the hole and electron spin polarization are shown in Figs.~\ref{QB}(b,c). The extracted parameters are: $\omega_{\rm L,e} = 0.120$~rad/ps,  $\omega_{\rm L,h} = 0.025$~rad/ps, and $T^*_{\rm 2,e} = 460$~ps, $T^*_{\rm 2,h} = 380$~ps at $B_\text{V} = 0.5$~T. The magnetic field dependences of the Larmor precession frequencies are plotted in Fig.~\ref{QB}(d). From linear fits with $\omega_{\rm L,e(h)}=|g_\text{V,e(h)}| \mu_{\rm B} B_{\rm V}/\hbar$, we obtain the electron and hole $g$-factor values: $|g_\text{V,e}| = 2.83$ and $|g_\text{V,h}| = 0.54$. The lack of a finite offset in the Zeeman splittings for $B_{\rm V}\to 0$ in Fig.~\ref{QB}(d) confirms that the signal originates from spatially separated electron-hole pairs with negligible exchange interaction. This interaction is expected to result in a finite offset for electron-hole pairs bound into excitons.

Taking into account that in MAPbI$_3$ the electron $g$-factor is positive and the hole $g$-factor is negative~\cite{kirstein2022nc,kirstein2022mapi}, we can calculate the bright exciton $g$-factor by $g_\text{V,X} = g_\text{V,h} + g_\text{V,e} = +2.29$, which is in good agreement with the value of $g_\text{F,X} = +2.4$ obtained from the Zeeman splitting of the PL emission line, measured in  longitudinal magnetic field (Fig.~\ref{DCP}(b)).

The spin dephasing times of electrons and holes evaluated from the decay in the $P_{\rm oo,e(h)}(t)$ dynamics (shown in Figs.~\ref{QB}(b,c)) are presented as function of magnetic field in Fig.~\ref{QB}(e). For the electrons, they can be reliably evaluated in the field range of $0.5-1$~T, while for the holes this can be done in the larger range of $0.5-3$~T. For both carrier types, the spin dephasing time decreases with increasing magnetic field in accordance with $1/B$ dependence. This is typical when the dephasing of a spin ensemble is provided by a dispersion of the carrier $g$-factor $\Delta g$~\cite{Spin_book_2017_Ch_6} due to inhomogeneity. 
%At $B_{\rm V}=0.5$~T, the times are $T^*_{\rm 2,e} = 460$~ps and $T^*_{\rm 2,h} = 380$~ps.

Note that commonly the coherent carrier spin dynamics are studied by time-resolved Faraday/Kerr rotation techniques~\cite{Spin_book_2017_Ch_6,kirstein2022am}, as we also did for MAPbI$_3$ crystals~\cite{kirstein2022mapi}. The method of time-resolved polarized photoluminescence, that we employ in the present study, was implemented in Ref.~\onlinecite{Heberle1994} for observation of spin quantum beats of carriers through the circular and linear polarization degrees of the emission from their recombination.

\subsection{Exciton spin polarization in longitudinal magnetic field}
\label{Exciton DCP}

\begin{figure*}[t!]
\begin{center}
\includegraphics[width = 17cm]{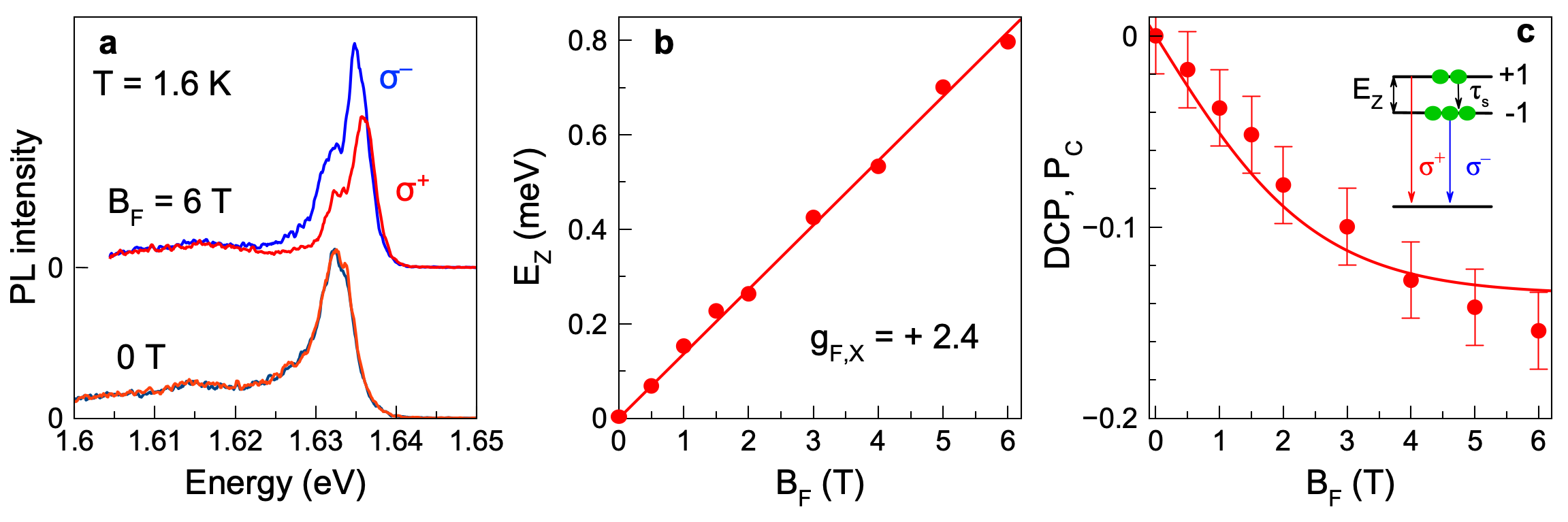}
\caption{\label{DCP} (a) PL spectra integrated over the time range $0-5$~ps after the exciting laser pulse, measured in $\sigma^+$ (red) and $\sigma^-$ (blue) polarization in the longitudinal magnetic fields of $B_\text{F} = 0$~T and 6~T. To avoid optical orientation, linearly polarized excitation is used with $E_\text{exc} = 1.771$~eV photon energy and $P = 10$~mW/cm$^2$ power density. $T = 1.6$~K. (b) Dependence of the exciton Zeeman splitting on $B_\text{F}$ (symbols). Linear fit (line) provides the exciton $g$-factor $g_{\text{F,X}} = +2.4$. (c) Dependence of the circular polarization degree induced by the Faraday field $B_\text{F}$ (symbols). A fit by Eq.~\eqref{eq2} with the only fitting parameter $\tau_\text{s,X} = 6\tau_\text{X} = 90$\,ps is shown by the red line. Insert is a scheme of the exciton thermalization on the Zeeman levels split by $B_\text{F}$.} 
\end{center}
\end{figure*}

To study the spin relaxation of cold excitons near the band gap, we measure the magnetic-field-induced degree of circular polarization (DCP) of the PL. In this experiment, a magnetic field is applied parallel to the light $k$-vector (Faraday geometry). The field causes a Zeeman splitting of the bright exciton states with spins $\pm1$ by $E_\text{Z} = g_\text{F,X} \mu_\text{B} B_\text{F}$. The exciton distribution is assumed to be thermalized on these spin levels (see scheme in Fig.~\ref{DCP}(c)), resulting in a polarization with the DCP degree described by 
\begin{equation}
\label{eqDCP}
P_\text{c}(B_\text{F}) = \frac{I^+ - I^-}{I^+ + I^-}.
\end{equation}
Here, $I^+$ and $I^-$ are the intensities of the $\sigma^+$ and $\sigma^-$ polarized emission, respectively. For that, the PL spectra are measured in $\sigma^+$ and $\sigma^-$ polarizations, while a linearly polarized excitation laser is used to exclude any contribution of optical orientation. 

PL spectra integrated over a temporal range of $0-5$~ps after the laser excitation pulse are shown in Fig.~\ref{DCP}(a) for $\sigma^+$ and $\sigma^-$ circular polarization. At zero magnetic field, the spectra are identical in the two polarizations, but at $B_\text{F} = 6$~T they are spectrally shifted from each other and have different intensities as result of polarization. The spectral separation of 0.8~meV between them corresponds to the exciton Zeeman splitting. The dependence of  $E_\text{Z}$ on  magnetic field is linear without zero-field offset, as shown in Fig.~\ref{DCP}(b). The red line represents a linear fit from which we evaluate the exciton $g$-factor $g_\text{F,X} = +2.4$. Note that the spectrum in $\sigma^-$ polarization has lower energy, which evidences the positive sign of $g_\text{F,X}$. The measured value of the exciton $g$-factor is in good agreement with the results of Ref.~\cite{Kopteva_gX_2024}, where it was shown for various bulk lead halide perovskites with band gaps in the visible spectral range of $1.52-3.21$~eV that the exciton $g$-factor varies from $+2.3$ to $+2.7$. 

We integrate the emission spectrum over the exciton recombination time of 15~ps and calculate the degree of circular polarization using Eq.~\eqref{eqDCP}. The DCP dependence on magnetic field is shown in Fig.~\ref{DCP}(c). The exciton thermalization on the Zeeman levels can be described by the equilibrium dependence~\cite{Parson1971}:
\begin{equation}
\label{eq2}
P_\text{c}(B_\text{F}) = - \frac{\tau_{\text{X}}}{\tau_{\text{X}} + \tau_{\text{s,X}}}\tanh \left(\frac{g_\text{F,X}\mu_\text{B}B_\text{F}}{2k_\text{B}T} \right).
\end{equation}
Here, $k_\text{B}$ is the Boltzmann constant. A fit with known from experiment parameters $\tau_\text{X} = 15$~ps and $T= 1.6$~K is presented in Fig.~\ref{DCP}(c) by the red line. It gives the exciton spin relaxation time of $\tau_\text{s,X} = 90$~ps as the only fit parameter.

\subsection{Modeling of polarized emission of excitons and carriers in longitudinal magnetic field}
\label{Modeling DCP}

It is important to analyze whether the magnetic-field-induced polarization of the exciton PL differs from that of the electron-hole recombination. If a difference is found, this can help to distinguish and separate these processes, even when their emission  spectrally overlap. In this Section, we perform a corresponding model analysis using the parameters of MAPbI$_3$. A complete investigation covering all possible parameter combinations is beyond the scope of this work and will be presented elsewhere.

The electron and hole polarizations in longitudinal magnetic field can be calculated as
\begin{equation}
\label{eq3}
P_\text{e(h)}(B_\text{F}) = - \frac{\tau_{\text{e(h)}}}{\tau_{\text{e(h)}} + \tau_{\text{s,e(h)}}}\tanh \left(\frac{g_\text{F,e(h)}\mu_\text{B}B_\text{F}}{2k_\text{B}T} \right).
\end{equation} 
Here $\tau_{\text{e(h)}}$ is the lifetime of electrons (holes) and $\tau_{\text{s,e(h)}}$ is the associated spin relaxation time. For the modeling, we use the parameters from experiment: $T=1.6$~K, $g_\text{e} = +2.8$, $g_\text{h} = -0.5$, and $g_\text{X} = g_\text{e} + g_\text{h}=+2.3$.

\begin{figure}[t]
\begin{center}
\includegraphics[width = 8.5cm]{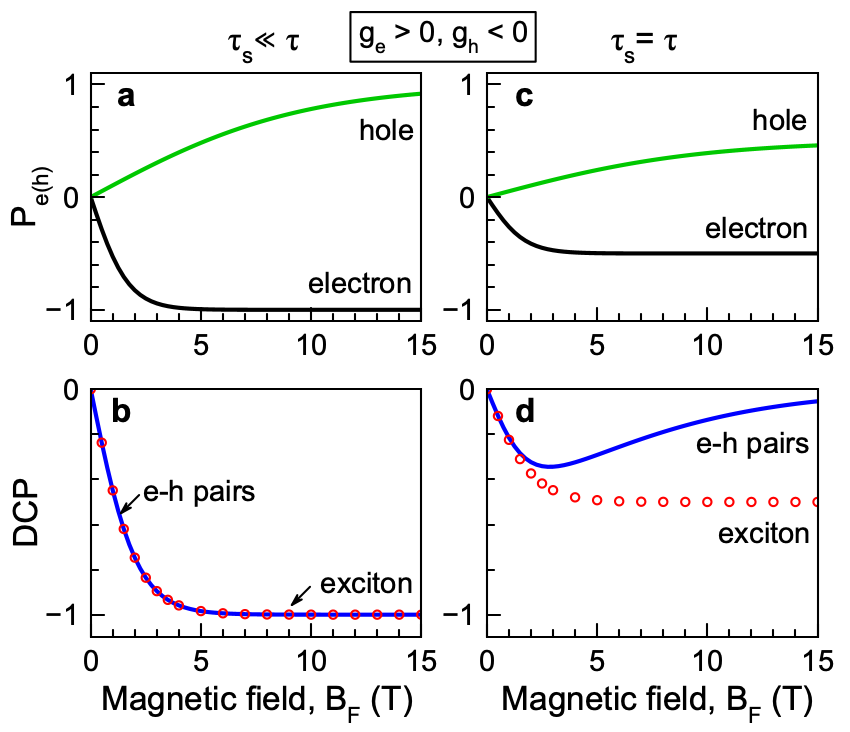}
\caption{\label{Sim_DCP} Calculation of the electron and hole polarization and the exciton and electron-hole DCP in a longitudinal magnetic field. (a) Case of a short spin relaxation time $\tau_{\text{s,e(h)}} \ll \tau_{\text{e(h)}}$ of electron (black) and hole (green) with $g_\text{e} = +2.8$ and $g_\text{h} = -0.5$. $T = 1.6$~K. (b) Blue line gives the circular polarization of electron-hole emission in dependence on the Faraday magnetic field $B_\text{F}$. Red symbols show polarization of exciton recombination, which are thermalized on the Zeeman levels for $\tau_{\text{s,X}} \ll \tau_{\text{X}}$. (c) Spin polarization of electrons (black) and holes (green) in case of $\tau_{\text{s,e(h)}} = \tau_{\text{e(h)}}$. (d) Blue line is the circular polarization of the electron-hole emission in dependence on $B_\text{F}$. Red symbols show polarization of exciton recombination, which are thermalized on the Zeeman levels  for $\tau_{\text{s,X}} = \tau_{\text{X}}$.}
\end{center}
\end{figure}

The electron and hole contributions are shown in Figs.~\ref{Sim_DCP}(a,c) for different ratios between the spin relaxation time and the carrier lifetime. In case of $\tau_{\text{s,e(h)}} \ll \tau_{\text{e(h)}}$, the prefactor in the right part of Eq.~\eqref{eq3} is 1, and the electron and hole polarizations reach in strong magnetic fields their maximal values of unity corresponding to 100\%, see Fig.~\ref{Sim_DCP}(a). For the holes with smaller $g$-factor, saturation on the maximal value occurs in stronger magnetic fields compared to the electrons. Also, the polarization signs for electrons and holes are opposite to each other, reflecting the opposite signs of their $g$-factors. For the case of $\tau_{\text{s,e(h)}} = \tau_{\text{e(h)}}$ shown in Fig.~\ref{Sim_DCP}(c), the prefactor equals to 0.5, which accordingly limits the maximum carrier polarization in high fields to 50\%. 

The DCP of the electron-hole recombination can be calculated by~\cite{kocher98_si}
\begin{equation}
\label{exciton:pol}
P_\text{c}(B_\text{F}) =\frac{P_\text{e}+P_\text{h}}{1+P_\text{e}P_\text{h}}.
\end{equation}
The DCP is shown by the blue line in Figs.~\ref{Sim_DCP}(b,d). Here the red circles show the exciton DCP calculated using Eq.~\eqref{eq2}. As shown in Fig.~\ref{Sim_DCP}(b), the exciton and e-h DCP are identical when $\tau_{\text{s}} \ll \tau$. However, for $\tau_{\text{s}} = \tau$, they are similar only in weak magnetic fields, but differ significantly in stronger fields, see Fig.~\ref{Sim_DCP}(d). In this case, the exciton DCP saturates, while the e-h DCP exhibits a nonmonotonic behavior. In this regime, the exciton and carrier recombination can be distinguished by the magnetic field dependences of DCP. Note that in a longitudinal magnetic field, bright exciton states are not mixed with spin-forbidden states from the singlet and triplet.

\section{Conclusions}
\label{Conclusions}

In conclusion, we have studied the spin dynamics of excitons and localized carriers in MAPbI$_3$ thin crystals with an orthorhombic crystal structure using the optical orientation technique. The spin polarization of the excitons and carriers is detected through the polarized photoluminescence from their recombination, and time-resolved measurements distinguish the origins of the different contributions. A high optical orientation degree is measured at cryogenic temperatures, reaching 85\% for excitons and 40\% for localized electrons and holes. This degree of orientation is remarkably robust with respect to a significant detuning of the excitation energy and is entirely suppressed for detunings exceeding 1.5 eV. We conclude that the Dyakonov-Perel spin relaxation mechanism is absent in orthorhombic MAPbI$_3$ crystals, which evidences that the spatial inversion symmetry is maintained.  The exciton spin polarization induced by a longitudinal magnetic field shows that the exciton spin relaxation time greatly exceeds the exciton lifetime. Larmor precession of the electron and hole spins is detected in a transverse magnetic field, and the carrier $g$-factors are evaluated. We also provide a model analysis of the optical orientation dynamics and the DCP for the case in which two distinguished spin systems contribute to the emission as is the case here. This situation is specific for perovskite semiconductors, where the exciton and carrier recombination overlap spectrally. This results in non-trivial spin dynamics and magnetic field dependences of the spin polarization and opens the way for obtaining an in-depth understanding of them. The combination of unusual spin properties, compared to conventional semiconductors, with the simple crystal fabrication and the bright optical properties establish lead halide perovskites as a promising platform for spintronic technologies.

\subsection*{Acknowledgements}
The authors are thankful for fruitful discussions to M. O. Nestoklon, M. M. Glazov, E. Kirstein, C. Harkort, and D. Kudlacik. We acknowledge the financial support by the Deutsche Forschungsgemeinschaft via the SPP2196 Priority Program (Projects YA 65/28-1, no. 527080192, and AK 40/13-1, no. 506623857). N.E.K. acknowledges the support of the Deutsche Forschungsgemeinschaft (project KO 7298/1-1, no. 552699366). The work at ETH Z\"urich (B.T., D.N.D., and M.V.K.) was financially supported by the Swiss National Science Foundation (grant agreement 200020E 217589) through the DFG-SNSF bilateral program and by ETH Z\"urich through ETH+ Project SynMatLab.

\section*{Author information}
\textbf{Nataliia E. Kopteva} -- orcid.org/0000-0003-0865-0393\\
\textbf{Dmitri R. Yakovlev} -- orcid.org/0000-0001-7349-2745\\
\textbf{Ey\"up~Yalcin} -- orcid.org/0000-0003-2891-4173\\
\textbf{Ilya~A. Akimov} -- orcid.org/0000-0002-2035-2324\\
\textbf{Mladen Kotur} -- orcid.org/0000-0002-2569-5051\\
\textbf{Bekir Turedi} -- orcid.org/0000-0003-2208-0737\\
\textbf{Dmitry N. Dirin} -- orcid.org/0000-0002-5187-4555 \\
\textbf{Maksym~V.~Kovalenko}  -- orcid.org/0000-0002-6396-8938 \\
\textbf{Manfred Bayer} -- orcid.org/0000-0002-0893-5949\\

\end{document}